\documentclass[aps,prd,reprint,nofootinbib,superscriptaddress]{revtex4-2}

\usepackage{mathrsfs}
\usepackage{amsfonts}
\usepackage{amsmath}
\usepackage{multirow}
\usepackage{slashed}
\usepackage{array}
\usepackage{verbatim}
\usepackage{graphicx}
\usepackage{tabularx}
\usepackage{bm}
\usepackage{ragged2e}
\usepackage{hyperref}
\usepackage{cleveref}

\begin{document}
\title{Redshift-Dependent Intrinsic Dispersion in the Quasar UV/X-ray Luminosity Relation}

\author{Jiaze Gao}
\email{gaojiaze@mail.dlut.edu.cn}
\affiliation{Institute of Theoretical Physics, School of Physics, Dalian University of Technology\\ Dalian 116024, People's Republic of China}

\author{Jianping Hu}
\affiliation{Ministry of Education Key Laboratory for Nonequilibrium Synthesis and Modulation of Condensed Matter, \\ School of Physics, Xi'an Jiaotong University, Xi'an 710049, People's Republic of China}
\affiliation{Key Laboratory of Modern Astronomy and Astrophysics (Nanjing University), Ministry of Education, \\ Nanjing 210093, People's Republic of China}

\author{Yun Chen}
\affiliation{National Astronomical Observatories, Chinese Academy of Sciences\\
Beijing 100101, People's Republic of China}
\affiliation{College of Astronomy and Space Sciences, University of Chinese Academy of Sciences\\
Beijing 100049, People's Republic of China}

\author{Bowen Zhao}
\affiliation{Institute of Theoretical Physics, School of Physics, Dalian University of Technology\\ Dalian 116024, People's Republic of China}

\author{Lixin Xu}
\email{lxxu@dlut.edu.cn}
\affiliation{Institute of Theoretical Physics, School of Physics, Dalian University of Technology\\ Dalian 116024, People's Republic of China}

\date{\today}

\begin{abstract}
Accurate modeling of the intrinsic dispersion in the quasar UV/X-ray luminosity relation is essential for reliable cosmological inference. We investigate its redshift dependence using luminosity distances reconstructed from cosmic chronometer and baryon acoustic oscillation measurements through Gaussian-process (GP) regression.
Bayesian model comparison and posterior constraints show that the intrinsic dispersion is not well described by a single redshift-independent constant over $0.7<z<2.6$. It remains approximately constant at $0.7<z<1.6$, but shows an overall decreasing trend in the higher-redshift interval $1.6<z<2.6$, where the redshift-dependent intrinsic-dispersion model is decisively favored. This conclusion remains qualitatively robust against changes in the scaling-relation parameterization, GP kernel, and redshift binning scheme.
We further examine its impact on cosmological inference in the flat $\Lambda$CDM model and find that, under the adopted calibration setup, the redshift-dependent intrinsic-dispersion model shifts the posterior median of $\Omega_{\rm m0}$ by $\Delta\Omega_{\rm m0}\simeq 0.025$. This indicates that intrinsic-dispersion modeling is a non-negligible component of the systematic-error budget for quasar cosmology and should be accounted for in future precision analyses.
\end{abstract}
\maketitle

\section{Introduction}

Quasars are among the most luminous active galactic nuclei (AGN). Their high luminosities and broad redshift distribution on cosmological scales make them promising candidates as standardizable candles, thereby extending the Hubble diagram into the redshift regime ($z \gtrsim 1.5$) where type Ia supernovae (SNe Ia) are scarce~\citep{Mortlock:2011va,Risaliti:2016nqt,Banados:2017unc,Lusso:2020pdb,Freedman:2021ahq,Riess:2020fzl,2021ApJ...907L...1W,Riess:2021jrx,Brout:2022vxf,2024NatAs...8..126B}.

Historically, extensive efforts have been devoted to employing quasars as standardizable candles~\citep{1977ApJ...214..679B,Osmer:1998mb,Elvis:2002ja,Watson:2011um,Wang:2013ha,Marziani:2013zra,LaFranca:2014eba,2015ApJ...801....8K}. Among the scaling relations investigated, the logarithmic linear relation between the ultraviolet (UV) luminosity at 2500\,\AA{} and the X-ray luminosity at 2\,keV has been the most widely studied and has been applied to several quasar samples~\citep{1979ApJ...234L...9T,1981ApJ...245..357Z,1986ApJ...305...83A,Risaliti:2015zla,Risaliti:2016nqt,2016ApJ...819..154L,Lusso:2020pdb,2022ApJ...931..106D,2022A&A...663L...7S,Lusso:2025bhy,2025A&A...703A.273S,2026A&A...706A.337L}. 

Physically, this luminosity relation is usually associated with the connection between the UV emission from the accretion disk and the X-ray emission from the corona~\citep{1994ApJ...436..599S,2002MNRAS.332..165M,2017A&A...602A..79L,2018MNRAS.480.1247K}. Although existing theoretical models still rely on simplifying assumptions or show partial inconsistencies with observations, they provide valuable entry points for studying disk physics with current and future large samples, and for developing quasars into precise cosmological probes~\citep{Risaliti:2016nqt,Lusso:2020pdb,2022A&A...663L...7S,2023A&A...676A.143S,Lusso:2025bhy,2025A&A...703A.273S,2026A&A...706A.337L}.

Analogous to SNe Ia, reducing and properly modeling the intrinsic dispersion beyond observational uncertainties is a prerequisite for developing quasars into high-precision standardizable candles~\citep{2011ApJ...740...72M,2014ApJ...793...16M,2016ApJ...819..154L,2016ApJ...822L..35S,2017ApJ...836...56K,2021ApJ...909...26B,Freedman:2021ahq,Riess:2020fzl,Riess:2021jrx,Brout:2022vxf}. Over the past decade, through improved analyses and increasingly stringent sample-cleaning procedures, the intrinsic dispersion $\delta_{\mathrm{int}}$ of large quasar samples suitable for cosmological applications has been reduced from an initial value of $\sim 0.4\,\mathrm{dex}$ to $\sim 0.2\,\mathrm{dex}$~\citep{Risaliti:2015zla,Risaliti:2016nqt,2016ApJ...819..154L,Lusso:2020pdb,Lusso:2025bhy,2025A&A...703A.273S}. 

Recently, a variety of approaches have been proposed to further reduce the intrinsic dispersion, including analyses of high-redshift subsamples~\citep{2022A&A...663L...7S}, the exploration of alternative observational proxies for accretion-disk and corona physics~\citep{2023A&A...676A.143S}, and the incorporation of additional physical information such as the photon index into the scaling relation~\citep{2026A&A...706A.337L}. At the same time, substantial progress has been made in understanding the origin of the intrinsic dispersion. Factors such as X-ray variability, accretion-disk inclination, and differences between spectroscopic and photometric flux measurements have been shown to contribute to the total dispersion~\citep{2019AN....340..267L,2023A&A...676A.143S,2024A&A...687A..32S,2024MNRAS.528.5972P,Prokhorenko:2025rbz}.

These studies have substantially improved our understanding of the physical origin and observational properties of the quasar UV/X-ray relation. Nevertheless, in most cosmological applications the intrinsic dispersion is still treated as a redshift-independent constant~\citep{2020PhRvD.102l3532Y,2022PhRvD.106d1301O,2024PDU....4401464O,2025PDU....4901975O,2020MNRAS.497..263K,2021MNRAS.502.6140K,2022MNRAS.517.1901L,2024MNRAS.530.4493Z,2025ChPhC..49g5101W,Lusso:2020pdb,Li:2024hed,2026arXiv260612265G}. This assumption is convenient, but it is not guaranteed: the physical properties of quasar populations, observational selection effects, and residual systematics may all vary with redshift. Several recent studies have also reported indications of redshift-dependent or non-universal dispersion in different quasar subsamples~\citep{2021MNRAS.502.6140K,2022MNRAS.517.1901L,Lusso:2020pdb,Lusso:2025bhy,2024A&A...687A..32S,Li:2024hed,2022A&A...663L...7S,2026arXiv260612265G}.

As quasars continue to develop into precise and reliable cosmological probes, it is important to characterize possible redshift dependence in the intrinsic dispersion and to assess the systematic effects that may arise if such dependence is neglected. In this paper, we use the quasar sample of \citet{Lusso:2020pdb}, together with luminosity distances reconstructed from cosmic chronometer (CC) and baryon acoustic oscillation (BAO) measurements through Gaussian-process (GP) regression, to investigate the intrinsic dispersion in the quasar UV/X-ray luminosity relation. We perform Bayesian model comparison between constant and redshift-dependent intrinsic-dispersion models, examine the robustness of our conclusions against several methodological choices, and further assess how the adopted intrinsic-dispersion model affects quasar-based cosmological parameter inference. We also place our findings in the context of previous studies of different quasar samples, additional observational proxies, and the dispersion budget, and discuss what they may imply for the possible origins of the intrinsic dispersion and for future quasar cosmology analyses.

The remainder of this paper is organized as follows: Sec.~\ref{sec:data and model} describes the quasar sample, the luminosity-distance reconstruction, and the Bayesian inference framework. Sec.~\ref{sec:result} presents the fiducial constraints on the redshift dependence of the intrinsic dispersion and the corresponding robustness tests. Sec.~\ref{sec:discussion} discusses the impact of intrinsic-dispersion modeling on cosmological parameter inference and compares our results with previous studies and possible physical or observational origins. Finally, Sec.~\ref{sec:summary} summarizes our conclusions.

\section{Data and Methodology}\label{sec:data and model}

\subsection{Quasar Sample and the UV/X-ray Relation}

We use the quasar sample compiled by \citet{Lusso:2020pdb} (L20). After a series of quality cuts designed to reduce known sources of systematic contamination, the sample contains approximately 2400 quasars, providing a large statistical basis for the present analysis. For each quasar, the sample provides the redshift $z$, the UV flux $F_{\mathrm{UV}}$ and its observational uncertainty $\sigma_{F_{\mathrm{UV}}}$, and the X-ray flux $F_{\mathrm{X}}$ and its observational uncertainty $\sigma_{F_{\mathrm{X}}}$.

An empirical log-linear scaling relation exists between the UV and X-ray luminosities of quasars, usually written as
\begin{equation}
\log L_{\mathrm{X}} = \gamma \log L_{\mathrm{UV}} + \beta,
\label{eq:LxLuv_relation}
\end{equation}
where $\log$ denotes the base-10 logarithm.

The luminosity $L$ and the observed flux $F$ are related through the definition of the luminosity distance $d_L$,
$F=\frac{L}{4\pi d_L^2}$. Substituting this relation into the above equation and taking logarithms transforms the luminosity relation into a direct relation between observable fluxes:
\begin{equation}
\log F_{\mathrm{X}} = \gamma \log F_{\mathrm{UV}} + (2\gamma - 2) \log d_L(z)
+ (\gamma - 1) \log(4\pi) + \beta.
\label{eq:Fx_relation}
\end{equation}

This relation connects the observed fluxes to the luminosity distance and therefore provides the basis for calibrating quasars as standardizable candles.

\subsection{Reconstructing Luminosity Distances}
\label{subsec:dl}

To calibrate the quasar UV/X-ray relation without assuming a specific parametric form of the expansion history, we reconstruct the luminosity distance $d_L(z)$ from CC and BAO-derived $H(z)$ measurements using GP regression.

The adopted $H(z)$ compilation, collected by \citet{2025MNRAS.542.1063H}, contains 60 measurements in the redshift range $0.07<z<2.40$: 35 CC measurements based on relative galaxy ages at different redshifts~\citep{Simon:2004tf,2012JCAP...08..006M,2014RAA....14.1221Z,2016JCAP...05..014M,2022ApJ...928L...4B,2015MNRAS.450L..16M,2017MNRAS.467.3239R,2023ApJS..265...48J,2023JCAP...11..047J,2023A&A...679A..96T,2002ApJ...573...37J,2022LRR....25....6M,2020ApJ...898...82M}, and 25 BAO-derived $H(z)$ measurements, which rely on the sound horizon at the drag epoch, including five recent DESI measurements~\citep{2025JCAP...02..021A,2025JCAP...01..124A,2025JCAP...04..012A,2009MNRAS.399.1663G,2017MNRAS.469.3762W,2017MNRAS.470.2617A,2013MNRAS.433.3559C,2014MNRAS.441...24A,2012MNRAS.425..405B,2020MNRAS.499..210N,2013A&A...552A..96B,2015A&A...574A..59D,2014JCAP...05..027F,2017A&A...608A.130D,2024MNRAS.534..544C,2024PhRvD.109b3525G}.

The reconstruction is performed in two steps.

\paragraph{Reconstruction of $H(z)$.}

Following the standard GP formalism, the Hubble parameter is modeled as
\begin{equation}
H(z)\sim \mathcal{GP}\bigl(\mu(z),k(z,z')\bigr),
\label{gp}
\end{equation}
where $\mu(z)$ and $k(z,z')$ denote the mean and covariance functions, respectively.

The kernel hyperparameters are determined by maximizing the marginal likelihood
\begin{equation}
\ln \mathcal{L} = -\frac{1}{2} \tilde{\mathbf{y}}^\top (K + \Sigma)^{-1} \tilde{\mathbf{y}} - \frac{1}{2} \ln \det(K + \Sigma) - \frac{n}{2} \ln 2\pi,
\label{gp_loglike}
\end{equation}
where $\tilde{\mathbf y}=\mathbf f-\mu(X)$ and $\Sigma$ is the observational covariance matrix.

For a set of prediction points $X_*$, the GP predictive mean and covariance are
\begin{equation}
\bar{\mathbf{f}}_* = \mu(X_*) + K(X_*,X) \bigl[ K(X,X) + \Sigma \bigr]^{-1} \tilde{\mathbf{y}},
\label{gp_mu}
\end{equation}
\begin{equation}
\small
\mathrm{Cov}(\mathbf{f}_*) = K(X_*,X_*) - K(X_*,X) \bigl[ K(X,X) + \Sigma \bigr]^{-1} K(X,X_*).
\label{gp_sigma}
\end{equation}

We use the CC and BAO measurements $\{z_i,H(z_i),\sigma_{H(z_i)}\}$ as the training set. To avoid imposing a specific cosmological expansion history, we adopt a constant mean function. Hyperparameter optimization is repeated from different initial conditions to avoid convergence to local maxima. The Gaussian-process reconstruction is performed using the publicly available \texttt{GaPP3} package~\citep{2012JCAP...06..036S}.

\begin{figure}[ht!]
\centering
\includegraphics[width=1.\linewidth]{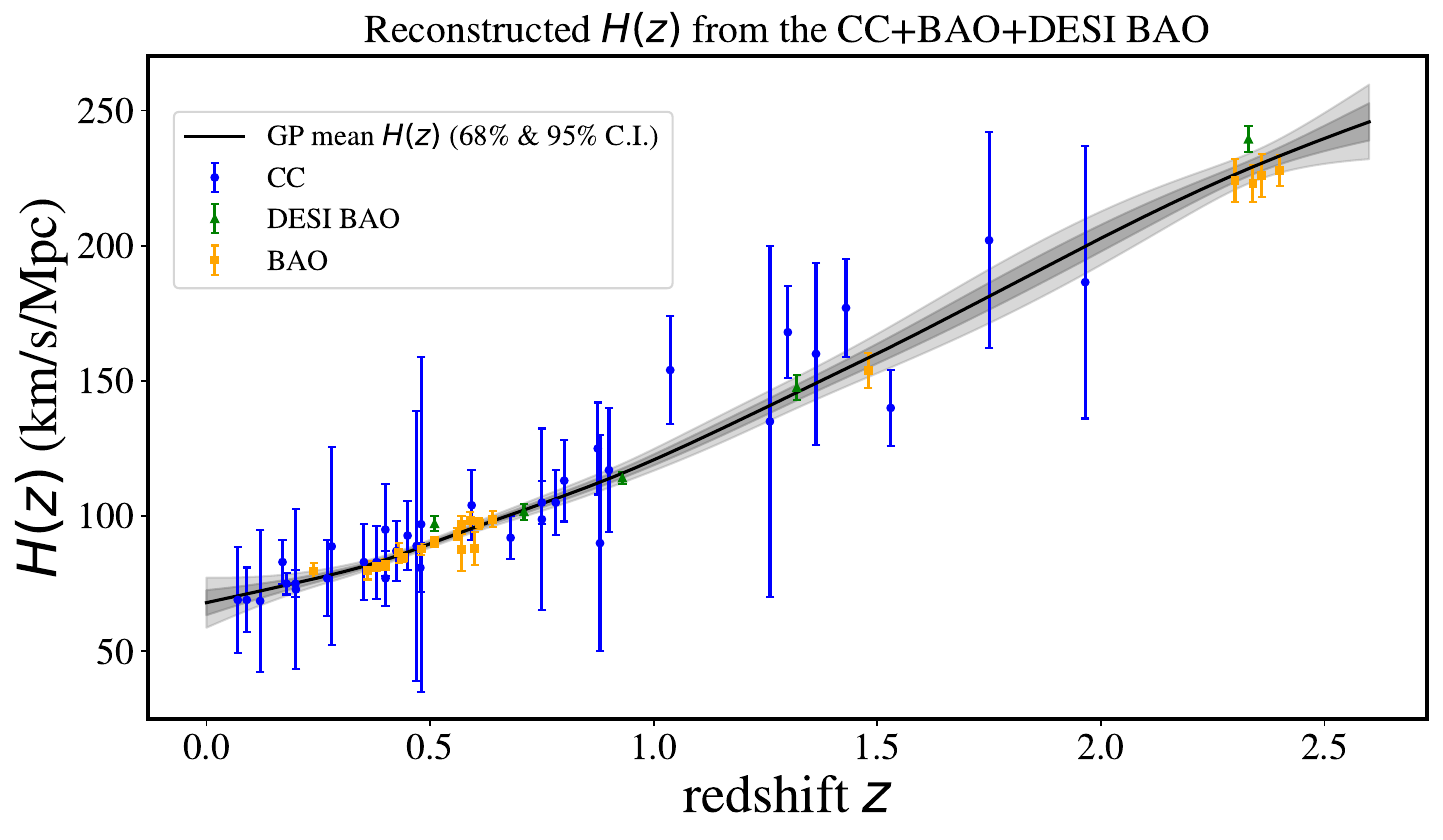}
\caption{Reconstructed Hubble parameter $H(z)$ from CC and BAO measurements. The blue circles, green triangles, and orange squares denote CC, DESI BAO, and other BAO measurements, respectively. The solid black curve represents the GP predictive mean, while the shaded regions indicate the 68\% and 95\% confidence intervals.}
\label{fig:hz}
\end{figure}

We adopt the Mat\'ern $3/2$ kernel as the fiducial choice and examine alternative kernels in the robustness tests. 
Fig.~\ref{fig:hz} shows the reconstructed Hubble parameter. The displayed uncertainty bands correspond to the diagonal elements of the GP covariance matrix, while the full covariance matrix is retained in all subsequent analyses.

\paragraph{Reconstruction of $d_L(z)$.}

On a dense redshift grid $\{z_j\}$, the GP reconstruction provides the predictive mean vector $\bar{\mathbf H}_*$ and covariance matrix $\mathrm{Cov}(\mathbf{H}_*)$. We draw 2000 realizations from the multivariate Gaussian distribution $\mathcal{N}\bigl(\bar{\mathbf{H}}_*, \mathrm{Cov}(\mathbf{H}_*)\bigr)$.

Assuming spatial flatness, $\Omega_K=0$, but without specifying a parametric form of $H(z)$, the luminosity distance for each realization is computed as
\begin{equation}
d_{L}(z) = c\,(1+z) \int_{0}^{z} \frac{\mathrm{d}z'}{H(z')},
\label{eq:dl}
\end{equation}
where the integral is evaluated numerically.

The ensemble of 2000 realizations yields the mean vector and covariance matrix of $d_L(z)$, which are adopted in the subsequent analysis. Since the lowest-redshift $H(z)$ measurement is located at $z\simeq0.07$, the integration involves a short GP extrapolation toward $z=0$. In addition, because the quasar analysis extends to $z=2.6$ whereas the adopted $H(z)$ compilation reaches $z\simeq2.40$, a modest high-redshift extrapolation is also involved. The associated uncertainties are propagated through the Monte Carlo realizations and included in the reconstructed covariance matrix. The robustness of these extrapolations is further examined empirically in the subsequent analysis by considering alternative GP kernels and by comparing the continuous redshift-dependent fit with the narrow-bin estimates.

\begin{figure}[ht!]
\centering
\includegraphics[width=1.\linewidth]{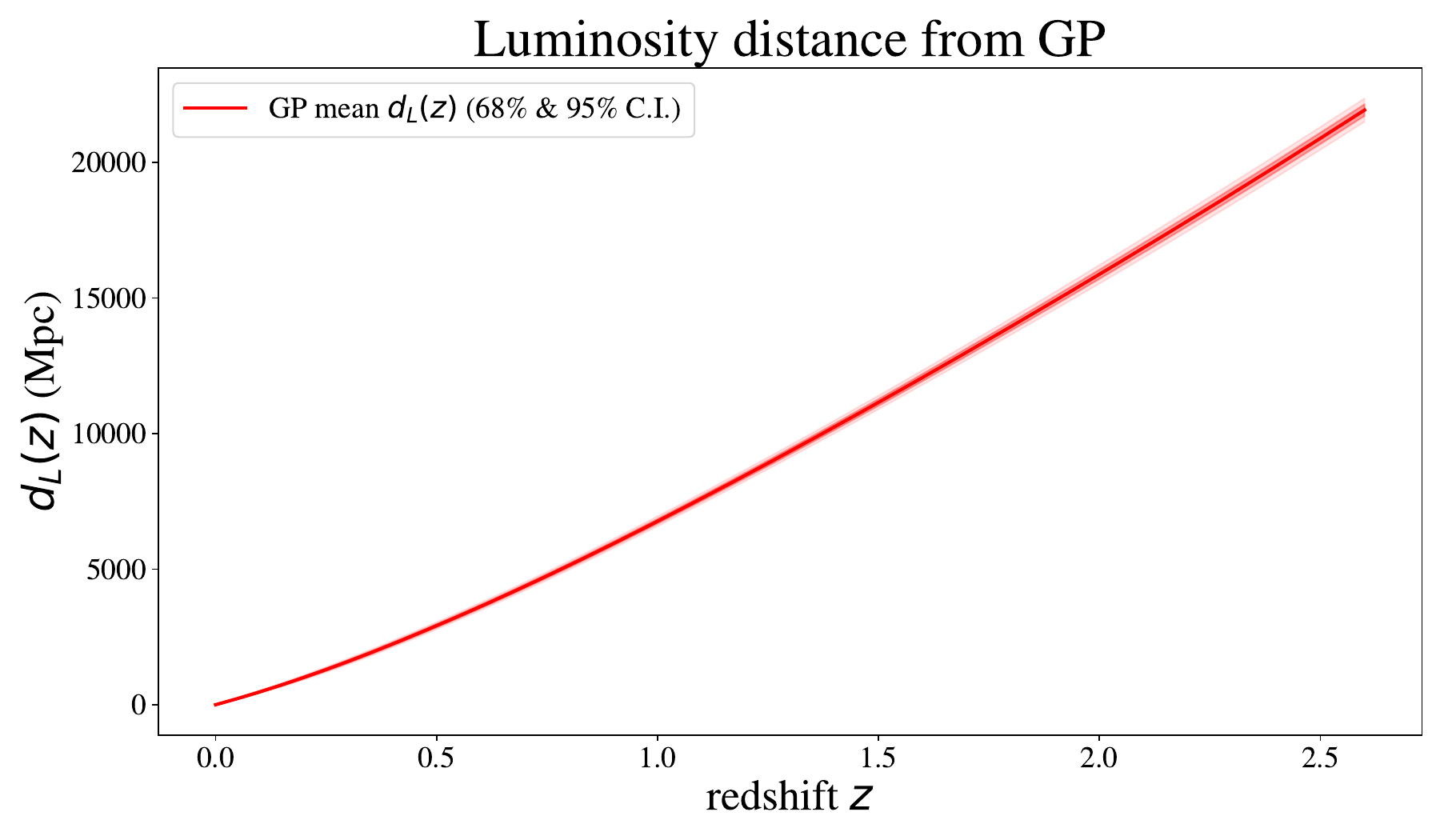}
\caption{Reconstructed luminosity distance $d_L(z)$ obtained from the GP reconstruction of $H(z)$. The solid curve denotes the ensemble mean of the Monte Carlo realizations, while the inner and outer shaded regions indicate the 68\% and 95\% confidence intervals, respectively.}
\label{fig:dl}
\end{figure}

Fig.~\ref{fig:dl} presents the reconstructed luminosity distance. The shaded regions indicate the marginal 68\% and 95\% confidence intervals derived from the reconstructed covariance matrix, while the full covariance matrix is used in parameter estimation.

\subsection{Bayesian Inference Framework}\label{bayes}

We adopt a Bayesian approach to infer the parameters of the quasar scaling relation and to compare different assumptions about the intrinsic dispersion through the Bayesian evidence. 

For a model $\mathcal{M}$ with parameters $\boldsymbol{\theta}$, Bayes' theorem reads
\begin{equation}
    p(\boldsymbol{\theta} \mid \mathcal{D},\mathcal{M}) = \frac{\mathcal{L}(\mathcal{D} \mid \boldsymbol{\theta},\mathcal{M}) \, p(\boldsymbol{\theta} \mid \mathcal{M})}{\mathcal{Z}(\mathcal{D} \mid \mathcal{M})},
    \label{eq:bayes}
\end{equation}
where $p(\boldsymbol{\theta} \mid \mathcal{M})$ is the prior distribution, $\mathcal{L}$ is the likelihood, and the Bayesian evidence $\mathcal{Z}$ is obtained by marginalizing the numerator over the full parameter space.

Given the reconstructed luminosity distances $d_L(z)$ and their covariance from Sec.~\ref{subsec:dl}, the theoretical log X-ray flux $\log F_{\mathrm{X,th}}$ for each quasar is computed from the scaling relation expressed in flux--flux form, as given by Eq.~(\ref{eq:Fx_relation}) for the classic relation. For the redshift-dependent scaling relation, the constant slope and intercept are replaced by $\gamma(z)=\gamma_0 + \frac{z}{1+z}\gamma_1$ and $\beta(z)=\beta_0 + \frac{z}{1+z}\beta_1$. We refer to this phenomenological extension as the $z$-correction relation.

Defining the residual vector between the observed and theoretical log fluxes as
\begin{equation}
    \Delta \mathbf{F} = \log \mathbf{F}_{\mathrm{X,obs}} - \log \mathbf{F}_{\mathrm{X,th}},
\end{equation}
and assuming that the residuals follow a multivariate Gaussian distribution, the corresponding log-likelihood is
\begin{equation}
    \ln \mathcal{L} = -\frac{1}{2} \Delta \mathbf{F}^{\mathrm{T}} \mathbf{C}^{-1} \Delta \mathbf{F} - \frac{1}{2} \ln \det \mathbf{C} - \frac{N}{2} \ln (2\pi),
    \label{eq:loglike}
\end{equation}
where $\mathbf{C}$ is the total covariance matrix and $N$ is the number of quasars.

The covariance matrix combines several sources of uncertainty. Its diagonal elements are
\begin{equation}
\small
    C_{ii} = \sigma_{\log F_{\mathrm{X}},i}^2 + \gamma_{\mathrm{eff}}(z_i)^2 \, \sigma_{\log F_{\mathrm{UV}},i}^2 + \delta_{\mathrm{int}}(z_i)^2 + g(z_i)^2 \, \sigma^2_{\log d_{L}(z_i)},
    \label{eq:Cdiag}
\end{equation}
with $\gamma_{\mathrm{eff}}(z)=\gamma$ for the classic model and $\gamma_{\mathrm{eff}}(z)=\gamma(z)$ for the $z$-dependent model, and $g(z) = 2\gamma_{\mathrm{eff}}(z)-2$. The term $\sigma^2_{\log d_{L}(z_i)}$ is the variance of $\log d_{L}(z_i)$ obtained from the GP reconstruction (Sec.~\ref{subsec:dl}). The intrinsic dispersion $\delta_{\mathrm{int}}(z)$ is either taken as a constant,
\[
\delta_{\mathrm{int}}(z) = \delta_{\mathrm{int}},
\]
or allowed to evolve with redshift as
\[
\delta_{\mathrm{int}}(z) = \delta_{\mathrm{int}} + \frac{z}{1+z}\,\delta_{\mathrm{int},1}.
\]

Since the observational flux uncertainties and intrinsic dispersion are assumed to be uncorrelated among different quasars, the off-diagonal elements arise only from the covariance of the reconstructed luminosity distances:
\begin{equation}
    C_{ij} = g(z_i)\, g(z_j) \, \mathrm{Cov}\bigl[\log d_{L}(z_i), \log d_{L}(z_j)\bigr], \qquad i \neq j.
    \label{eq:Coff}
\end{equation}

To isolate the effect of intrinsic-dispersion evolution while allowing for possible redshift dependence of the scaling relation itself, we consider four model configurations. For the classic UV/X-ray relation, we fit either a constant intrinsic dispersion, with parameters $(\gamma,\beta,\delta_{\mathrm{int}})$, or a redshift-dependent intrinsic dispersion, with parameters $(\gamma,\beta,\delta_{\mathrm{int}},\delta_{\mathrm{int},1})$. For the $z$-correction relation, we similarly consider a constant-dispersion model with parameters $(\gamma_0,\beta_0,\gamma_1,\beta_1,\delta_{\mathrm{int}})$ and an evolving-dispersion model with parameters $(\gamma_0,\beta_0,\gamma_1,\beta_1,\delta_{\mathrm{int}},\delta_{\mathrm{int},1})$.

For the prior distribution $p(\boldsymbol{\theta} \mid \mathcal{M})$, we adopt uniform priors
$\gamma,\gamma_0\in[0,1]$,
$\beta,\beta_0\in[0,15]$,
$\delta_{\rm int}\in[0,3]$,
and $\gamma_1,\beta_1,\delta_{\rm int,1}\in[-2,2]$.

We sample the posterior distributions and evaluate the Bayesian evidence $\mathcal{Z}$ with the nested-sampling code \texttt{PyMultiNest}~\citep{Skilling:2004pqw,10.1111/j.1365-2966.2009.14548.x,Buchner:2014nha}, and visualize the posterior distributions using \texttt{getdist}~\citep{2019arXiv191013970L}. For all parameters, we report the median and 68\% credible intervals of their marginal posterior distributions.

In the following tables, parameter constraints are rounded according to their posterior uncertainties and to the precision needed for the subsequent comparisons. Most uncertainties are reported with two significant digits, while selected parameters are kept to three decimal places when a finer comparison is required.

\section{Redshift Dependence of the Intrinsic Dispersion}\label{sec:result}
\subsection{Fiducial Results}\label{main_result}

Previous studies based on redshift-binned quasar subsamples have suggested that the intrinsic dispersion may be smaller at higher redshifts, and possible redshift dependence of the scaling relation has been reported around $z\sim 1.5$--$1.7$~\citep{Lusso:2020pdb,Lusso:2025bhy,2022MNRAS.517.1901L,2026arXiv260612265G,Li:2024hed}. At the low-redshift end, sources with $z<0.7$ may be affected by residual contamination, and have been reported to show behavior different from that of the higher-redshift quasar sample in scaling-relation analyses~\citep{Lusso:2020pdb,Lusso:2025bhy,2026arXiv260612265G,Li:2024hed}. We therefore focus on the redshift range $0.7<z<2.6$. Motivated by the above indications, we split this redshift range into three intervals: a low-redshift interval $0.7<z<1.6$, a high-redshift interval $1.6<z<2.6$, and the full redshift range considered in this work, $0.7<z<2.6$.

For each scaling-relation parameterization and each redshift interval, we compute the posterior distributions of the model parameters for the constant intrinsic-dispersion model and for the redshift-dependent intrinsic-dispersion model. We also list the corresponding logarithmic Bayesian evidence $\ln \mathcal{Z}$ and the logarithmic evidence of the evolution model relative to the constant model, $\Delta \ln \mathcal{Z}=\ln \mathcal{Z}_\mathrm{evolution}-\ln \mathcal{Z}_\mathrm{constant}$. These results are summarized in Table~\ref{tab:results}. Throughout this section, the strength of Bayesian evidence is interpreted using the empirical scale of \citet{Kass:1995loi} and \citet{Trotta:2008qt}: $0 \le \Delta \ln \mathcal{Z}<1$ (inconclusive), $1 \le \Delta \ln \mathcal{Z}<3$ (positive), $3 \le \Delta \ln \mathcal{Z}<5$ (strong), and $\Delta \ln \mathcal{Z} \ge 5$ (decisive).

\begin{table*}[htbp]
    \renewcommand{\arraystretch}{1.1}
    \setlength{\tabcolsep}{5pt}
    \centering
    \caption{Posterior constraints on the scaling-relation and intrinsic dispersion parameters for different redshift intervals and model configurations.}\label{tab:results}
    \begin{tabular}{cccccccccc}
        \hline
        Scaling model&$\delta_{\mathrm{int}}$ model&$\gamma$ & $\beta$ &$\delta_{\mathrm{int}}$&$\gamma_1$ & $\beta_1$  &$\delta_{\mathrm{int},1}$&  $\ln\mathcal{Z}$ & $\Delta \ln \mathcal{Z}$\\
        \hline
        \hline
        \multicolumn{10}{c}{$0.7<z<1.6$}\\
        \hline
        \multirow{2}{*}{classic relation} & constant & $0.60 ^{+0.02}_{-0.02}$ & $8.18 ^{+0.46}_{-0.47}$ & $0.234 ^{+0.005}_{-0.005}$ & $-$ & $-$ & $-$ & $0.59$ & $0$ \\
        & evolution & $0.60 ^{+0.02}_{-0.02}$ & $8.17 ^{+0.47}_{-0.49}$ & $0.31 ^{+0.05}_{-0.04}$ & $-$ & $-$ & $-0.14 ^{+0.08}_{-0.09}$ & $-0.57$ & $-1.16$ \\
        \hline
        \multirow{2}{*}{$z$-correction relation} & constant & $0.58 ^{+0.03}_{-0.03}$ & $8.71 ^{+0.80}_{-0.82}$ & $0.234 ^{+0.005}_{-0.005}$ & $0.01 ^{+0.04}_{-0.05}$ & $0.01^{+1.37}_{-1.36}
$ & $-$ & $-3.03$ & $0$ \\
        & evolution & $0.58 ^{+0.03}_{-0.03}$ & $8.74  ^{+0.84}_{-0.84}$ & $0.31^{+0.05}_{-0.05}$ & $0.01 ^{+0.04}_{-0.05}$ & $-0.07^{+1.46}_{-1.31}$ & $-0.14^{+0.09}_{-0.09}$ & $-4.03$ & $-1.00$ \\
        \hline
        \hline
        \multicolumn{10}{c}{$1.6<z<2.6$}\\
        \hline
        \multirow{2}{*}{classic relation} & constant & $0.53 ^{+0.02}_{-0.02}$ & $10.48 ^{+0.59}_{-0.59}$ & $0.207 ^{+0.006}_{-0.007}$ & $-$ & $-$ & $-$ & $43.18$ & $0$ \\
        & evolution & $0.53 ^{+0.02}_{-0.02}$ & $10.51^{+0.54}_{-0.56}$ & $0.79 ^{+0.15}_{-0.14}$ & $-$ & $-$ & $-0.87^{+0.21}_{-0.23}$ & $49.71$ & $6.53$ \\
        \hline
        \multirow{2}{*}{$z$-correction relation} & constant & $0.50^{+0.03}_{-0.03}$ & $10.81^{+0.97}_{-0.99}$ & $0.205^{+0.007}_{-0.006}$ & $0.03^{+0.05}_{-0.04}$ & $0.03^{+1.32}_{-1.37}$ & $-$ & $43.30$ & $0$ \\
        & evolution & $0.50^{+0.03}_{-0.03}$ & $10.92^{+1.00}_{-1.02}$ & $0.83^{+0.15}_{-0.14}$ & $0.03^{+0.04}_{-0.04}$ & $0.00^{+1.37}_{-1.39}$ & $-0.95^{+0.21}_{-0.22}$ & $51.58$ & $8.28$ \\
        \hline
        \hline
        \multicolumn{10}{c}{$0.7<z<2.6$}\\
        \hline
        \multirow{2}{*}{classic relation} & constant & $0.62 ^{+0.01}_{-0.01}$ & $7.64^{+0.34}_{-0.34}$ & $0.230^{+0.004}_{-0.004}$ & $-$ & $-$ & $-$ & $18.00$ & $0$ \\
        & evolution & $0.62^{+0.01}_{-0.01}$ & $7.67^{+0.33}_{-0.34}$ & $0.33^{+0.03}_{-0.03}$ & $-$ & $-$ & $-0.17^{+0.05}_{-0.05}$ & $20.15$ & $2.15$ \\
        \hline
        \multirow{2}{*}{$z$-correction relation} & constant & $0.55^{+0.03}_{-0.03}$ & $9.54^{+0.81}_{-0.83}$ & $0.225^{+0.004}_{-0.004}$ & $0.02^{+0.05}_{-0.04}$ & $0.10^{+1.33}_{-1.39}$ & $-$ & $45.35$ & $0$ \\
        
        & evolution & $0.54^{+0.03}_{-0.03}$ & $9.61^{+0.90}_{-0.80}$ & $0.35^{+0.03}_{-0.03}$ & $0.02^{+0.05}_{-0.04}$ & $0.14^{+1.30}_{-1.44}$ & $-0.23^{+0.05}_{-0.05}$ & $52.26$ & $6.90$ \\
        \hline
    \end{tabular}
\vspace{2mm} \begin{minipage}{0.98\textwidth} \footnotesize \justifying
Posterior constraints for the constant and redshift-dependent intrinsic-dispersion models in different redshift intervals. Results are shown for both the classic UV/X-ray relation and the $z$-correction relation. For the redshift-dependent model, the parameter $\delta_{\mathrm{int}}$ denotes the intercept of the adopted linear parameterization in $z/(1+z)$, while the effective intrinsic dispersion within the fitted redshift range is given by $\delta_{\mathrm{int}}(z)$. The column $\ln\mathcal{Z}$ gives the Bayesian evidence for each model, while $\Delta \ln \mathcal{Z}$ denotes the evidence difference relative to the corresponding constant-dispersion model within the same redshift interval and scaling-relation parameterization.
\end{minipage}
\end{table*}

\begin{figure*}[ht!]
\centering  \includegraphics[width=0.95\linewidth]{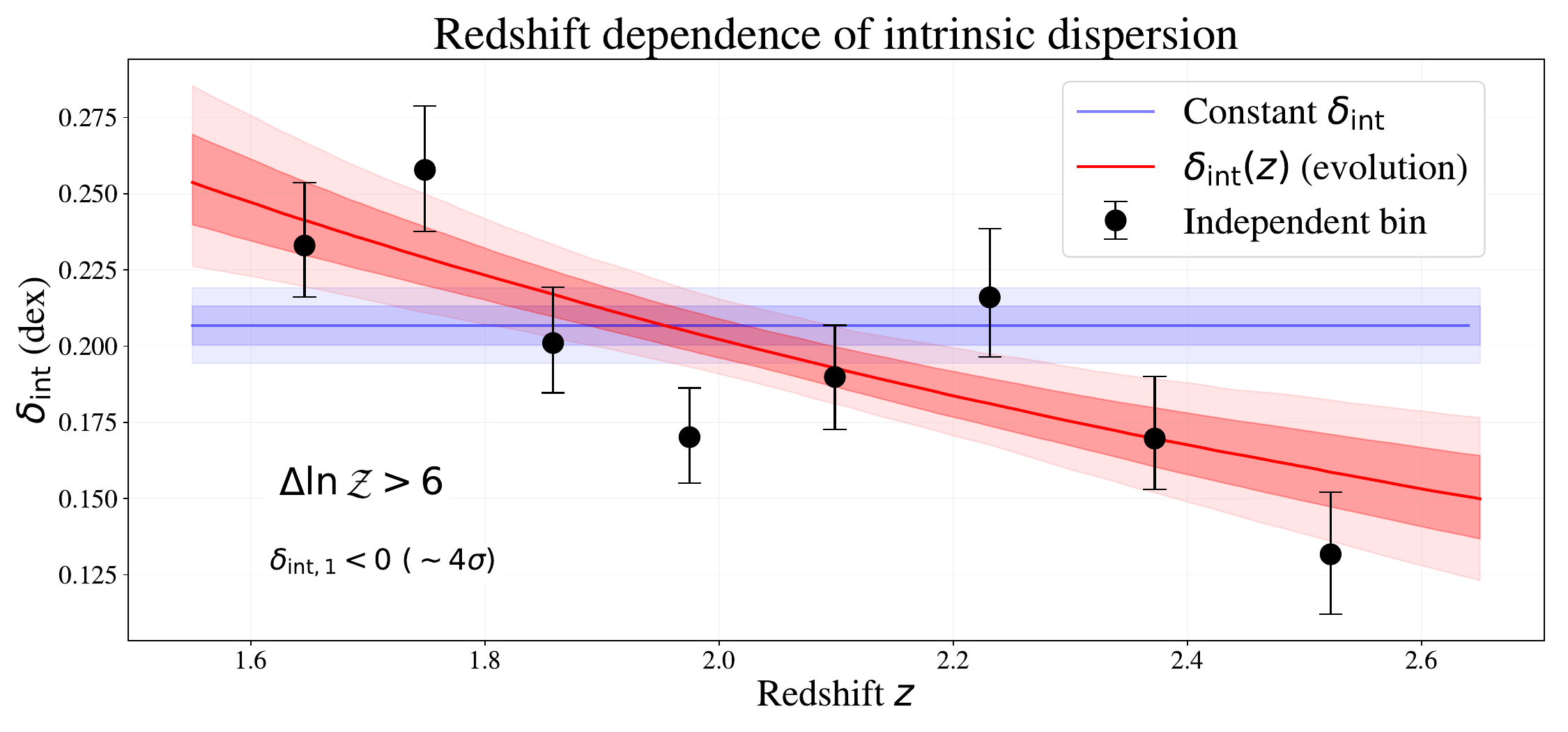}
	\caption{Evolution of the intrinsic dispersion in the quasar UV/X-ray luminosity relation. 
The red solid curve shows the posterior mean of the evolving intrinsic-dispersion model, with shaded regions indicating the 68\% and 95\% credible intervals. 
The blue dashed line corresponds to the conventional constant-dispersion model. 
Black points represent bin-by-bin estimates obtained independently of the continuous redshift-dependent fit. 
For the high-redshift interval $1.6<z<2.6$ under the classic relation, the evolving-dispersion model is favored with $\Delta\ln\mathcal{Z}=6.53$, and $\delta_{\rm int,1}=0$ is disfavored at approximately $4\sigma$ within this parameterization.
}
    \label{fig:z_delta}
\end{figure*}
We begin with the classic scaling relation, which provides the simplest case for assessing the redshift dependence of the intrinsic dispersion.
In the redshift interval $0.7 < z < 1.6$, the intrinsic dispersion evolution parameter $\delta_{\mathrm{int},1}$ is consistent with zero within $\sim 1.7\sigma$, and the Bayesian evidence difference of the evolution model relative to the constant model is $\Delta \ln \mathcal{Z} = -1.16$, which slightly favors the constant intrinsic-dispersion model.

In the interval $1.6 < z < 2.6$, $\delta_{\rm int,1}=0$ is disfavored at $\sim 4\sigma$ within this parameterization, and the relative Bayesian evidence reaches $\Delta \ln \mathcal{Z} = 6.53$, corresponding to decisive Bayesian evidence in favor of the redshift-dependent intrinsic-dispersion model over the constant-dispersion model.

In Fig.~\ref{fig:z_delta}, we compare the bin-by-bin intrinsic dispersion values obtained from narrow-bin fitting with the prediction of the continuous redshift-dependent model. The bin-by-bin estimates broadly follow the same qualitative trend as the redshift-dependent model, while also indicating possible local deviations. The Bayesian evidence further quantifies the statistical preference for this model over the constant-dispersion case.

Under the classic relation with a constant dispersion fitted separately in the two redshift intervals, the inferred values of $\delta_{\rm int}$ differ at the level of about $3\sigma$. This result is qualitatively compatible with the systematically smaller dispersion values inferred from high-redshift subsamples in previous studies~\citep{Lusso:2020pdb,Lusso:2025bhy,2022MNRAS.517.1901L,2026arXiv260612265G,Li:2024hed}. 

Over the full range $0.7<z<2.6$, $\delta_{\rm int,1}=0$ is disfavored at $\sim 3\sigma$, while the relative Bayesian evidence, $\Delta\ln\mathcal{Z}=2.15$, provides positive support for the redshift-dependent intrinsic-dispersion model. 

The global analysis is therefore consistent with the two-interval picture: the effective intrinsic dispersion is approximately constant at lower redshift and tends to be smaller at higher redshift. This is also consistent with the result that fitting a constant dispersion separately in two broad redshift bins yields a lower value in the high-redshift bin. Such a value can be interpreted as an average over an effectively decreasing dispersion within that interval. Overall, these results indicate that the intrinsic dispersion is not adequately described by a single redshift-independent constant.

We cannot yet draw a firm conclusion on whether the scaling-relation parameters and the intrinsic-dispersion parameter evolve in a coordinated manner. It is, however, worth noting that both exhibit a change in behavior beyond $z \sim 1.5-1.7$. From a physical perspective, both the scaling-relation parameters and the intrinsic dispersion are jointly determined by quasar physics and data selection effects. Such a coincidence may reflect changes in the underlying quasar population, redshift-dependent selection effects, or a combination of both.

\subsection{Robustness Tests}

The preference for a redshift-dependent intrinsic dispersion may in principle be affected by several methodological choices. In this subsection, we examine three of them: the parameterization of the quasar UV/X-ray scaling relation, the choice of GP kernel used for the luminosity-distance reconstruction, and the adopted redshift interval.

We first investigate whether our conclusions depend on the assumed form of the quasar UV/X-ray relation. In addition to the classic scaling relation, we consider a redshift-dependent parameterization in which both the slope and intercept are allowed to evolve with redshift (see Sec.~\ref{bayes} for details). After introducing these additional degrees of freedom, the main results remain qualitatively unchanged. The intrinsic dispersion is consistent with being approximately constant in the range $0.7<z<1.6$, while a decreasing trend is still preferred in the higher-redshift interval $1.6<z<2.6$. The primary difference is that the redshift-dependent scaling relation is statistically favored over the classic relation when the full sample is considered, leading to stronger evidence for a redshift-dependent intrinsic-dispersion model. This indicates that the preference for an evolving intrinsic dispersion is not simply an artifact of enforcing a redshift-independent scaling relation.

Since the inferred intrinsic dispersion could in principle be affected by the luminosity-distance reconstruction, we further test the impact of the GP kernel by replacing the fiducial Mat\'ern $3/2$ kernel with both the Mat\'ern $5/2$ and squared-exponential (RBF) kernels. For each kernel, the full reconstruction and parameter-estimation procedure is repeated under both forms of the scaling relation. The preference for a decreasing intrinsic dispersion at higher redshift persists for all kernel choices, indicating that the result is not driven by a specific GP covariance function.

A further concern is whether the inferred redshift dependence is sensitive to the adopted redshift boundaries. The fiducial division at $z=1.6$ was motivated by the narrow-bin analysis presented above, but alternative binning schemes may lead to different statistical conclusions if the underlying evolution is more complex than the simple linear model assumed here. To investigate this possibility, we perform additional analyses using four alternative redshift intervals: $1.4<z<2.6$, $1.8<z<2.6$, $1.6<z<2.4$, and $1.6<z<2.5$, considering both forms of the scaling relation.

The results reveal a broadly consistent picture. Three of the four alternative intervals continue to favor the redshift-dependent intrinsic-dispersion model, with Bayesian evidence ranging from positive to decisive and with $\delta_{\rm int,1}$ differing from zero at approximately $3$--$4\sigma$ significance. The only exception is the $1.8<z<2.6$ interval, for which the evidence mildly favors the constant-dispersion model. This exception is informative rather than contradictory. As indicated by the bin-by-bin estimates in Fig.~\ref{fig:z_delta}, the intrinsic dispersion does not appear to decrease monotonically throughout this narrower high-redshift range; instead, a local flattening or increase around $z\sim2.2$ weakens the evidence for a simple monotonic evolution. Therefore, the result does not rule out redshift-dependent intrinsic dispersion, but suggests that its detailed redshift dependence may be more complex than the single linear form adopted here.

Overall, these tests indicate that the main conclusion is not driven by any particular choice of scaling-relation parameterization, GP kernel, or redshift binning scheme. The broad agreement between the narrow-bin estimates and the continuous redshift-dependent fit suggests that the adopted linear form in $z/(1+z)$ captures the dominant trend present in the current data. Nevertheless, this parameterization should be regarded as a phenomenological and representative description rather than a unique functional form. The variation of the evidence among different redshift intervals, especially the weaker preference for evolution in the $1.8<z<2.6$ interval, indicates that the intrinsic-dispersion evolution may contain local structure beyond a simple linear redshift dependence.

\section{Implications and Discussion}\label{sec:discussion}
\subsection{Sensitivity of Cosmological Parameters to the Intrinsic-dispersion Model}
\label{sec:cosmology}

To use quasars as reliable cosmological probes, it is essential to understand how assumptions about the intrinsic-dispersion model propagate into cosmological parameter inference. Since both the UV/X-ray scaling relation and the intrinsic dispersion may exhibit redshift-dependent behavior, the goal of this subsection is not to obtain competitive quasar-only cosmological constraints. Instead, we perform a controlled test designed to isolate the impact of the intrinsic-dispersion model itself.

For this purpose, we fix the scaling-relation parameters to the values obtained from the classic-relation fit in the high-redshift interval in Sec.~\ref{sec:result}, and vary only the intrinsic-dispersion model. This choice should not be interpreted as assuming that the scaling relation is fully stable. Rather, it allows us to separate the effect of intrinsic-dispersion modeling from possible changes in the slope and intercept of the scaling relation. It may also be viewed as an illustrative limit in which the quasar scaling relation has been independently calibrated, so that the remaining question is how the adopted intrinsic-dispersion model affects the inferred cosmological parameters.

All tests are performed in the redshift interval $1.6<z<2.6$, where the evidence for redshift-dependent intrinsic dispersion is strongest. Restricting the analysis to this interval allows us to examine the impact of the intrinsic-dispersion model in the regime where this effect is most pronounced.

We adopt the classic scaling relation and fix the scaling-relation parameters to $\gamma=0.53$ and $\beta=10.5$. We impose a Gaussian prior on the Hubble constant based on the SH0ES measurement, $H_0 = 73.04 \pm 1.04\,{\rm km~s^{-1}~Mpc^{-1}}$~\citep{Riess:2021jrx}. We otherwise use the same likelihood and parameter-estimation procedure as described in Sec.~\ref{bayes}. The intrinsic dispersion is treated either as a redshift-dependent function or as a constant. The luminosity distances are computed in two cosmological models: a flat $\Lambda$CDM model, with $\Omega_{\rm m0}$ as the main cosmological parameter constrained by the quasar data and $H_0$ calibrated by the external prior, and the CPL model, with $\Omega_{\rm m0}$, $w_0$, and $w_a$ as free parameters.

\begin{figure}[ht!]
\centering
\includegraphics[width=1.\linewidth]{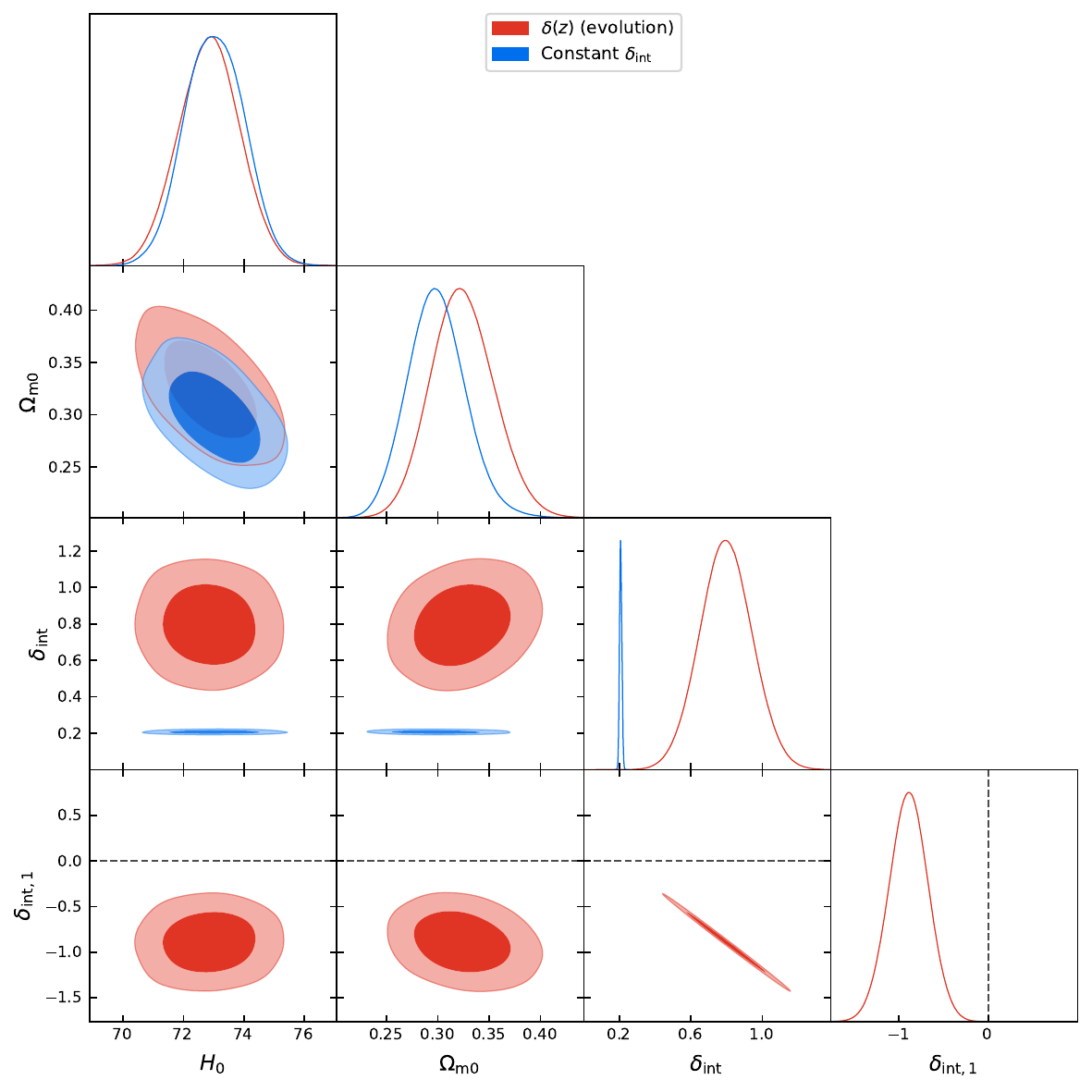}
\caption{
Posterior distributions obtained under the constant and redshift-dependent intrinsic-dispersion models in the redshift interval $1.6<z<2.6$ for the flat $\Lambda$CDM case with the SH0ES $H_0$ prior.
The inferred value of $\Omega_{\rm m0}$ shifts from $0.298^{+0.028}_{-0.027}$ to $0.323^{+0.032}_{-0.030}$ when the redshift-dependent intrinsic-dispersion model is adopted.
Contours correspond to the 68\% and 95\% credible regions.
The black dashed line indicates $\delta_{\mathrm{int},1}=0$.
}
\label{fig:delta_triangle}
\end{figure}

The resulting parameter constraints are summarized in Table~\ref{table:cosmos}, while the posterior distributions under the $\Lambda$CDM model are shown in Fig.~\ref{fig:delta_triangle}.

For the flat $\Lambda$CDM model, the constant-dispersion and evolving-dispersion assumptions yield $\Omega_{\rm m0}=0.298^{+0.028}_{-0.027}$ and $\Omega_{\rm m0}=0.323^{+0.032}_{-0.030}$, respectively. The shift in the posterior median is therefore $\Delta\Omega_{\rm m0}\simeq0.025$. Since the scaling-relation parameters are fixed in this controlled test, this shift should not be interpreted as the full systematic bias in a complete quasar cosmological analysis. Instead, it quantifies how changing only the intrinsic-dispersion model can move the inferred cosmological parameters under otherwise fixed assumptions. The Bayesian evidence favors the evolving-dispersion model, with $\Delta\ln\mathcal{Z}=7.21$, indicating that the preference for redshift-dependent intrinsic dispersion persists in this cosmological test.

For the CPL model, the constant-dispersion assumption gives $\Omega_{\rm m0}=0.393^{+0.044}_{-0.195}$, $w_0=-1.95^{+0.54}_{-0.65}$, and $w_a=-0.31^{+4.22}_{-3.17}$. The corresponding constraints under the evolving-dispersion model are $\Omega_{\rm m0}=0.421^{+0.035}_{-0.062}$, $w_0=-2.12^{+0.68}_{-0.58}$, and $w_a=-0.87^{+3.95}_{-2.81}$. The CPL parameters remain weakly constrained by the quasar sample, and the shifts in $w_0$ and $w_a$ are smaller than their posterior uncertainties. Nevertheless, the same qualitative behavior is observed: changing the intrinsic-dispersion model leads to visible shifts in the posterior distribution, while the evolving-dispersion model is favored by the Bayesian evidence, with $\Delta\ln\mathcal{Z}=6.71.$

We further test whether this conclusion depends on the treatment of $H_0$. As robustness checks, we repeat the analysis under two alternative treatments of $H_0$: fixing $H_0=73.04\,{\rm km~s^{-1}~Mpc^{-1}}$ and imposing a Gaussian prior centered on the Planck-inferred value, $H_0=67.4\pm0.5\,{\rm km~s^{-1}~Mpc^{-1}}$~\citep{2020A&A...641A...6P}. The absolute value of $\Omega_{\rm m0}$ changes, as expected, because of the correlation between $H_0$ and $\Omega_{\rm m0}$ in the distance relation. However, the relative impact of the intrinsic-dispersion model remains. With the Planck-like prior, the flat $\Lambda$CDM constraints shift from $\Omega_{\rm m0}=0.408^{+0.031}_{-0.033}$ for the constant-dispersion model to $\Omega_{\rm m0}=0.428^{+0.034}_{-0.032}$ for the evolving-dispersion model, with $\Delta\ln\mathcal{Z}=6.67.$ The CPL case shows the same qualitative behavior. These tests indicate that the sensitivity of cosmological inference to the intrinsic-dispersion model is not driven by a particular choice of the $H_0$ prior.

It is important to emphasize that the shifts induced by the evolving intrinsic-dispersion model are much smaller than the large discrepancies reported in some quasar-only cosmological analyses when compared with standard probes such as CMB, SNe Ia, and BAO measurements~\citep{2020PhRvD.102l3532Y,2022PhRvD.106d1301O,2024PDU....4401464O,2025PDU....4901975O,2020MNRAS.497..263K,2021MNRAS.502.6140K}. Therefore, redshift-dependent intrinsic dispersion should not be regarded as a complete explanation of the quasar cosmological anomaly. Rather, the present test shows that the intrinsic-dispersion model is one component of the systematic-error budget that can propagate into cosmological parameter inference.

The present exercise therefore illustrates that assumptions about the intrinsic-dispersion model can produce visible shifts in cosmological parameter inference, even when the scaling-relation parameters are held fixed. A more complete assessment will require flexible modeling of $\delta_{\rm int}(z)$ together with simultaneous calibration of the UV/X-ray scaling relation. Such analyses will be necessary for quantifying the impact of redshift-dependent intrinsic dispersion on previous quasar standardization studies and on future precision cosmological applications of quasars.

\begin{table*}[htbp]
        \renewcommand{\arraystretch}{1.1}
        \setlength{\tabcolsep}{6pt}
        \centering
	\caption{Comparison of cosmological parameter constraints obtained under constant and evolving intrinsic-dispersion models. The parameter $H_0$ is quoted in units of ${\rm km~s^{-1}~Mpc^{-1}}$.}\label{table:cosmos}
		\begin{tabular}{cccccccccc}
			\hline
			 Cosmology&$\delta_{\mathrm{int}}$ model&$H_0$&$\Omega_{\mathrm{m0}}$ &$w_0$&$w_a$&$\delta_{\mathrm{int}}$&$\delta_{\mathrm{int},1}$&  $\ln \mathcal{Z}$ & $\Delta \ln \mathcal{Z}$\\
			\hline
			\multirow{2}{*}{$\Lambda$CDM}&constant&$73.03^{+0.99}_{-0.98}$& $0.298^{+0.028}_{-0.027}$& $-$& $-$ &$0.21^{+0.01}_{-0.01}$  & $-$ & $50.08$& $0$ \\
			&evolution& $72.88^{+0.98}_{-1.05}$&$0.323^{+0.032}_{-0.030}$& $-$& $-$ &$0.79^{+0.15}_{-0.14}$ & $-0.89^{+0.21}_{-0.21}$ & $57.29$& $7.21$ \\
			\hline
			\multirow{2}{*}{CPL}&constant&$72.98^{+0.92}_{-0.95}$ & $0.393^{+0.044}_{-0.195}$&$-1.95^{+0.54}_{-0.65}$&  $-0.31^{+4.22}_{-3.17}$&$0.21^{+0.01}_{-0.01}$& $-$ & $51.18$& $0$ \\
			&evolution&$72.95^{+0.88}_{-0.97}$&$0.421^{+0.035}_{-0.062}$& $-2.12^{+0.68}_{-0.58}$&  $-0.87^{+3.95}_{-2.81}$&$0.80^{+0.13}_{-0.14}$ & $-0.90^{+0.20}_{-0.19}$ & $57.89$& $6.71$ \\
			\hline
	  \end{tabular}
\vspace{2mm} \begin{minipage}{0.98\textwidth} \footnotesize
\justifying
Constraints on $H_0$, $\Omega_{\rm m0}$, $w_0$, and $w_a$
derived using quasars in the redshift range $1.6<z<2.6$
with fixed scaling-relation parameters ($\gamma=0.53$, $\beta=10.5$).
A Gaussian SH0ES prior is imposed on $H_0$.
The purpose of this comparison is not to derive competitive cosmological constraints,
but to evaluate the sensitivity of cosmological inference to the adopted intrinsic-dispersion model.
Two intrinsic-dispersion models, constant and redshift-dependent, are considered for both $\Lambda$CDM and CPL cosmologies.
The evolving-dispersion model is decisively favored by the Bayesian evidence and yields measurable shifts in the inferred cosmological parameters.
Quoted errors correspond to 68\% credible intervals.
\end{minipage}
\end{table*}

\subsection{Comparison with Previous Studies and Possible Origins}\label{subsec:source}

The goal of this subsection is not to identify a unique physical origin of the inferred redshift dependence, but to place our result in the context of previous measurements and known contributions to the dispersion budget.

Previous studies have reported a wide range of intrinsic-dispersion measurements for the quasar UV/X-ray relation. In this subsection, we compare our results with those obtained from different datasets and alternative band proxies, and discuss several possible astrophysical and observational mechanisms that may contribute to the observed redshift dependence.

First, although the L20 sample remains the benchmark dataset for quasar cosmology, its intrinsic dispersion is still substantially larger than that of traditional cosmological probes, indicating that further improvements in sample selection and physical understanding are required before quasars can reach their full cosmological potential. To this end, we qualitatively review the intrinsic dispersion in different historical datasets.

After correcting for systematic errors, the \citet{Risaliti:2015zla} sample reduced its mean intrinsic dispersion to $\sim0.3\,\mathrm{dex}$. This sample did not show a simple monotonic decrease of the dispersion over the full high-redshift range, although lower dispersion values were found in part of the intermediate-redshift interval. The binning study of \citet{Lusso:2020pdb} shows a decrease of intrinsic dispersion in the range $1.6<z<2.6$, consistent with our conclusion, while at $z>2.6$ the dispersion rises back to around the global mean. The origin of this behavior remains unclear and warrants further investigation. In addition, \citet{2025A&A...703A.273S} used early eROSITA (eFEDS) data to construct a quasar sample extending to $z\sim3$, and likewise found a significantly reduced intrinsic dispersion in the redshift bin centered at $\langle z\rangle\approx2.1$. The analyses of high-redshift subsamples suggest that at $z>2.5$ the intrinsic dispersion can be as low as $\sim0.12\,\mathrm{dex}$, and even $\sim0.07\,\mathrm{dex}$ at $z>3$~\citep{2022A&A...663L...7S}. Although this appears to support a systematic decrease of intrinsic dispersion in high-redshift quasars, the conclusion remains to be tested due to differences in sample selection and limited sample sizes---it may reflect the small dispersion of high-quality samples rather than a direct effect of redshift.

Second, from the perspective of different band proxies, \citet{2023A&A...676A.143S} employed other bands as proxies for the accretion disk and corona physics. When using UV spectral energy flux instead of luminosity flux, the difference in intrinsic dispersion between low and high redshift bins persists. When using 1\,keV flux as a proxy for X-ray flux, the intrinsic dispersion begins to decrease significantly from the bin with $\langle z\rangle\approx1.15$ and remains low up to high redshifts, without a clear linear decline. When using the 1\,keV X-ray flux and the Mg\,II $\lambda$2800\,\AA\ emission line as alternative proxies, a trend of decreasing intrinsic dispersion with increasing redshift is still observable.

In addition to sample selection and the choice of observational proxies, the dispersion budget is also relevant for interpreting the redshift dependence found in this work. \citet{2024A&A...687A..32S} decomposed the intrinsic dispersion into several components and found that accretion-disk inclination contributes $\sim 0.06\,\mathrm{dex}$, luminosity contributes $\sim 0.02\,\mathrm{dex}$, while X-ray variability provides the largest contribution, about $0.08\,\mathrm{dex}$. Their luminosity-binned analysis further showed that the variability contribution reaches $\sim 0.11\,\mathrm{dex}$ for low-luminosity quasars ($\log L_{\rm X}<26.9$), compared with only $\sim 0.02\,\mathrm{dex}$ for high-luminosity systems.

However, this variability budget should not be directly transferred to the L20 cosmological sample without accounting for differences in sample composition. Using the same luminosity threshold, we find that the ratio of low- to high-luminosity quasars in the L20 sample over $0.7<z<2.6$ is approximately $3:1$, whereas the corresponding subsamples in the variability-budget analysis are close to a $1:1$ ratio. Since the X-ray variability contribution depends strongly on luminosity, a different luminosity composition can change the effective contribution of variability to the total intrinsic dispersion. Moreover, the L20 sample shows a clear luminosity--redshift coupling, with higher-redshift quasars being, on average, more luminous. This suggests that part of the apparent redshift dependence of the intrinsic dispersion may be related to the changing luminosity composition of the sample, although a quantitative decomposition is beyond the scope of the present work.

Relatedly, \citet{2024MNRAS.528.5972P} showed that the X-ray variability amplitude on a fixed timescale is anti-correlated with black-hole mass, Eddington ratio, and luminosity. Differences in luminosity distributions, observational time baselines, black-hole masses, and Eddington-ratio distributions among redshift bins may therefore also contribute to the observed behavior of the intrinsic dispersion. A dedicated analysis that jointly accounts for luminosity, variability, and X-ray observing history will be needed to determine how much of the redshift dependence can be explained by these effects.

Taken together, variations in the intrinsic dispersion over specific redshift intervals can be seen across several quasar samples and related analyses. From the perspective of the currently identified dispersion budget, X-ray variability is an important contributor and may account for part of the observed behavior. However, its quantitative contribution cannot be directly transferred between samples without accounting for differences in luminosity distribution, observing cadence, and sample selection. If the identified components are still insufficient to explain the behavior seen in large cosmological quasar samples, additional physical or observational effects may need to be considered, including redshift-dependent quasar population properties and residual selection effects. These factors should also be considered in the construction and validation of more homogeneous quasar samples for cosmological applications, because they may affect not only the redshift dependence of the intrinsic dispersion but also that of the scaling relation, which enters cosmological parameter inference more directly. Future deep and homogeneous quasar surveys, together with multi-epoch X-ray observations, will be important for distinguishing among these possibilities and clarifying the origin of the observed redshift dependence.

\section{Summary}\label{sec:summary}
In this paper, we used luminosity distances reconstructed from CC and BAO measurements through GP regression to investigate the redshift dependence of the intrinsic dispersion in the quasar UV/X-ray relation. The main conclusions are as follows.

(i) The intrinsic dispersion of the quasar UV/X-ray relation is not adequately described by a single redshift-independent constant over the full redshift range considered. It remains approximately constant at $0.7<z<1.6$, but tends to be smaller at higher redshift. In the interval $1.6<z<2.6$, $\delta_{\rm int,1}=0$ is disfavored at approximately $4\sigma$ within the adopted parameterization, and the redshift-dependent intrinsic-dispersion model is decisively favored by the Bayesian evidence ($\Delta\ln\mathcal{Z}>6$).

(ii) This conclusion is not driven by the assumed form of the UV/X-ray scaling relation. Allowing the slope and intercept to evolve with redshift does not remove the preference for a redshift-dependent intrinsic dispersion. Additional tests with alternative GP kernels and redshift intervals further indicate that the main trend is robust, although the strength of the evidence depends on the chosen redshift range.

(iii) The intrinsic-dispersion model can affect quasar-based cosmological inference. In the redshift interval where the evidence for redshift dependence is strongest, replacing the conventional constant-dispersion treatment with the redshift-dependent intrinsic-dispersion model shifts the posterior median of $\Omega_{\rm m0}$ by $\Delta\Omega_{\rm m0}\simeq0.025$ in the flat $\Lambda$CDM model with a SH0ES prior on $H_0$. This result should be interpreted as a controlled sensitivity test rather than as a complete estimate of the systematic bias in quasar cosmology, but it shows that intrinsic-dispersion modeling can propagate into cosmological parameter inference.

(iv) Variations in the inferred intrinsic dispersion over specific redshift intervals can be seen across previous quasar samples and related analyses, especially in selected high-redshift or high-quality subsamples. X-ray variability, luminosity dependence, population differences, and redshift-dependent selection effects may all contribute to this behavior. These effects should be considered in future modeling of both the intrinsic dispersion and the UV/X-ray scaling relation, as well as in the construction of more homogeneous quasar cosmology samples.

These results suggest that redshift-dependent intrinsic dispersion should be taken into account in future precision applications of quasars as cosmological probes. Larger, more homogeneous quasar samples and multi-epoch X-ray observations will be important for clarifying its origin and quantifying its impact on cosmological analyses.

\begin{acknowledgments}
We thank Ziyu Guo for valuable discussions. This work has been supported by the National Key Research and Development Program of China (No. 2022YFA1602903), and the National Natural Science Foundation of China (Nos.~12473002 and 12475047).
\end{acknowledgments}

\section*{Data and Software Availability}
The quasar sample is taken from the publicly available catalog of \citet{Lusso:2020pdb}. The $H(z)$ data used for the luminosity-distance reconstruction are from the compilation of \citet{2025MNRAS.542.1063H} and the references therein. The analysis was performed using \texttt{GaPP3}, \texttt{PyMultiNest}, and \texttt{getdist}. The derived data products and analysis scripts are available from the corresponding author upon reasonable request.

\bibliographystyle{apsrev4-2}
\bibliography{bibfile}
\end{document}